# Deep Learning-based QSAR Model for Therapeutic Strategies Targeting SmTGR Protein's Immune Modulating Role in Host-Parasite Interaction

Belaguppa Manjunath Ashwin Desai
School of Engineering
Dayananda Sagar University
Bengaluru, India
ashwin-ece@dsu.edu.in

Belaguppa Manjunath Anirudh
Khoury College of Computer Sciences
Northeastern University
Boston, USA
anirudhdesai777@gmail.com

Kalyani.S.Biju
School of Basic and Applied Sciences
Dayananda Sagar University
Bengaluru, India
kalyanishree040@gmail.com

Vondhana Ramesh
School of Basic and Applied Sciences
Dayananda Sagar University
Bengaluru, India
vonds98@gmail.com

Pronama Biswas*
School of Basic and Applied Sciences
Dayananda Sagar University
Bengaluru, India
pronama-sbas@dsu.edu.in

*Abstract*——**Schistosomiasis, a neglected tropical disease caused by *Schistosoma* parasites, remains a major global health challenge. The *Schistosoma mansoni* thioredoxin glutathione reductase (SmTGR) is essential for parasite redox balance and immune evasion, making it a key therapeutic target. This study employs predictive Quantitative Structure-Activity Relationship (QSAR) modeling to identify potential SmTGR inhibitors. Using deep learning, a robust QSAR model was developed and validated, achieving high predictive accuracy. The predicted novel inhibitors were further validated through molecular docking studies, which demonstrated strong binding affinities, with the highest docking score of -10.76 ± 0.01 kcal/mol. Visualization of the docked structures in both 2D and 3D confirmed similar interactions for the inhibitors and commercial drugs, further supporting their therapeutic effectiveness and the predictive ability of the model. This study demonstrates the potential of QSAR modeling in accelerating drug discovery, offering a promising avenue for developing novel therapeutics targeting SmTGR to improve schistosomiasis treatment.**

*Keywords— Computer Aided Drug Discovery; Deep Learning; QSAR; Schistosoma mansoni Thioredoxine Glutathionine Reductase (SmTGR); Molecular docking.*

## I. INTRODUCTION

Schistosomiasis, or bilharzia, is a major global health concern, affecting millions in sub-Saharan Africa, the Caribbean, South America, Southeast Asia, and the Middle East, and causing around 280,000 deaths annually [1]. It is caused by *Schistosoma* species, primarily *S. haematobium, S. japonicum*, and *S. mansoni*. The parasites require freshwater snails as intermediate hosts before infecting humans, leading to chronic infections with abdominal pain, diarrhea, and organ damage [2].

*Schistosoma mansoni* Thioredoxin Glutathione Reductase (SmTGR) is a multifunctional enzyme essential for the parasite's survival in its mammalian host. It is a fusion of Thioredoxin Reductase (TR) and a redox-active Glutaredoxin (Grx) domain at its N-terminus. SmTGR plays a critical role in thiol redox metabolism by detoxifying reactive oxygen species (ROS) and maintaining redox balance. Since *S. mansoni* lacks typical mammalian glutathione and thioredoxin systems, it depends entirely on SmTGR. Identified as a key drug target; SmTGR offers a new therapeutic approach, especially since praziquantel, the sole schistosomiasis treatment, shows frequent failures and does not prevent reinfection [3].

To address the limitations of the existing therapeutic strategy, a quantitative structure-activity relationship (QSAR) model based on deep learning was developed and used to identify potential novel inhibitors of SmTGR [4]. Molecular docking studies were conducted to explore the interactions

This work was funded by the Karnataka State Council for Science and Technology (KSCST): 47S_MSC_0054

between these predicted compounds and SmTGR. This combined QSAR and docking approach highlights potential drug candidates for schistosomiasis therapy by targeting SmTGR.

## II. MATERIALS AND METHODS

### A. Data Collection and preprocessing

The ChEMBL database (https://www.ebi.ac.uk/chembl/) was used to identify potential inhibitors and retrieve data on their inhibitory activities. The target selected for this study, identified by ChEMBL-ID "CHEMBL6110," was chosen due to its extensive "potency" data for *Schistosoma mansoni*, encompassing a diverse range of compounds and assay conditions. The dataset was specifically filtered for potency values and compiled into a DataFrame, which was then saved as "Rawdata_TGR.csv" containing 28487 molecules. Subsequently, the dataset underwent a cleaning process to ensure data quality. Entries with missing potency values, invalid measurements (such as values denoted by "<" or ">"), and those lacking SMILES representations were removed. Potency values were standardized to nanomolar (nM) units. Redundant compounds were assessed for consistency, and those exhibiting high standard deviations in potency values were excluded. After these refinements, the final curated dataset, consisting of 28257 molecules, was saved as "Cleaned_TGR.csv."

### B. Descriptors and Fingerprint Calculation

Using RDKit (https://www.rdkit.org/), 210 molecular descriptors were computed from SMILES strings, capturing essential physicochemical and structural properties. Invalid molecules were identified and removed to ensure data accuracy. The final set of descriptors was then saved as "RDkit_Descriptors_TGR.csv". In addition, various molecular fingerprints including Morgan, Avalon, MACCS keys, topological torsion, and atom pair fingerprints were generated to assess structural characteristics. These fingerprints were combined into a single dataset and saved as "RDkit_FPs_TGR.csv" for further modeling. Together, the molecular descriptors and fingerprints contributed a total of 2937 features for QSAR modeling.

### C. Data Preprocessing, Feature Selection and Scaling

Potency values were transformed into pPotency and classified into two categories: active (pPotency $\geq$ 5) and inactive (pPotency < 5). To ensure data quality, all columns were examined for missing or infinite values, and only those with finite potency values were retained. The final dataset comprised 5569 active compounds (Class 1) and 22688 inactive compounds (Class 0).

Feature selection was performed by first removing low-variance features using a variance threshold, followed by mutual information-based filtering to retain relevant features (447). All non-binary features were scaled using StandardScaler from Scikit-learn. The pre-processed dataset was saved as "Classification_data_TGR.csv."

### D. Model Training and Testing

The dataset used in this study comprises molecular descriptors for compounds classified based on their inhibitory activity. To address the class imbalance, the minority class was oversampled using the resample method in Scikit-learn, ensuring a balanced representation of both active and inactive compounds. The dataset was then split into training (80%) and test (20%) sets using a stratified sampling approach to maintain class distribution. A deep feedforward neural network was designed for classification, consisting of four fully connected layers. The input layer accepted molecular features, followed by three hidden layers with 512, 256, and 128 neurons, respectively, each employing ReLU activation, batch normalization, and dropout regularization (0.4). The output layer contained a single neuron with a sigmoid activation function for binary classification. Kaiming normal initialization was used for weight initialization to improve convergence. The model was implemented using PyTorch.

The training process was conducted using Binary Cross Entropy (BCE) Loss, optimized with the Adam optimizer (learning rate = 0.001). The model was trained for 75 epochs with a batch size of 64, following a structured approach: during each epoch, the input data was passed through the network, the loss was computed between predictions and actual values, and weight updates were performed using backpropagation and gradient descent. To prevent overfitting, dropout regularization, and batch normalization were incorporated into the network. To ensure reproducibility, the entire workflow was executed with a fixed random seed (42), and the trained model was saved for prediction.

### E. Tools and Libraries

This study was conducted using Python, leveraging a range of specialized libraries for molecular descriptor generation, machine learning, deep learning, and data visualization. RDKit was employed for molecular descriptor and fingerprint generation, while Scikit-learn (https://scikit-learn.org) facilitated data pre-processing, model development, and performance evaluation. Deep learning experiments were implemented using PyTorch (https://pytorch.org) for neural network training, with NumPy and Pandas utilized for efficient numerical computations and data handling. Data visualization was carried out using Matplotlib (https://matplotlib.org) and Seaborn for graphical representation of results, including ROC curves and confusion matrices.

Computational tasks were executed on a high-performance server equipped with an AMD EPYC 7742 64-core processor, 503 GiB of RAM, and NVIDIA A100 GPUs, running on Ubuntu 22.04.4. All processes, including dataset preparation, model training, and evaluation, were conducted within a dedicated Conda environment to ensure reproducibility and dependency management.

### F. Dataset for Prediction

To discover novel SmTGR inhibitors, a dataset consisting exclusively of Genuine Natural (GN) and Naturally Sourced (NS) compounds was obtained from the MolPort database (https://www.molport.com). GN compounds occur naturally but can also be synthesized, whereas NS compounds are directly derived from natural sources. This dataset was analyzed using a trained model to predict the inhibitory activity of these compounds against SmTGR. This strategy enabled the identification of potential inhibitors with significant biological relevance. Moreover, since these compounds fall under the GN or NS categories, they are expected to exhibit lower toxicity or enhanced biocompatibility [4].

### G. Molecular Docking Methods

The predicted inhibitors were validated through molecular docking studies [5]. The binding affinities were compared to that of Digoxin (a known inhibitor of SmTGR) (CID: 2724385) [6]. The protein structure of SmTGR was retrieved as a Protein Data Bank (PDB) file, which was obtained from RCSB PDB when searched with the PDB ID "2V6O". The quality of the protein was assessed using PROCHECK (https://saves.mbi.ucla.edu/) and VoroMQA (https://bioinformatics.lt/wtsam/voromqa) web tools. The stereochemistry of the SMILES structures was corrected before conversion to SDF format. These SDF files were then converted to PDBQT format with Gasteiger charges added using an Open Babel command. Digoxin and SmTGR were prepared using AutoDockTools-1.5.7 (https://autodock.scripps.edu/) and saved in PDBQT format. Molecular docking was conducted using a Vina Script-based approach within a Conda environment on a server, accessed via Visual Studio Code. The predicted molecules were docked in batches using the same script-based method. The protein-ligand docked complex was examined to visualize amino acids and binding pockets using UCSF Chimera 1.17 (https://www.cgl.ucsf.edu/chimera/) and Discovery Studio 2021Client (https://www.3ds.com/products/biovia/discovery-studio/visualization). Fig. 1 shows the summary of the entire methodology followed in this study.

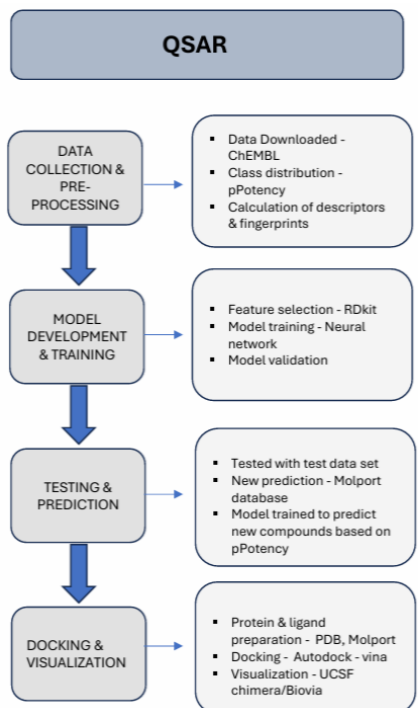

Fig. 1. Schematic representation of the QSAR modeling, molecular docking, and visualization methodology involved in the study. The figure was made using Microsoft® PowerPoint® 2021 MSO (Version 2501 Build 16.0.18429.20132).

## III. RESULTS AND DISCUSSIONS

### A. Dataset Characteristics

Following data cleaning and pre-processing, the dataset comprised 28,257 molecules, including 22,688 inactive compounds (pPotency < 5) and 5,569 active ones (pPotency ≥ 5). Fig. 2 illustrates the distribution of pPotency values across both active and inactive molecules. As shown, most inactive compounds are clustered around a pPotency value of 4.5, whereas active compounds display a more even distribution, particularly between pPotency values of 5 and 7. The class imbalance was handled using the oversampling of the minority class.

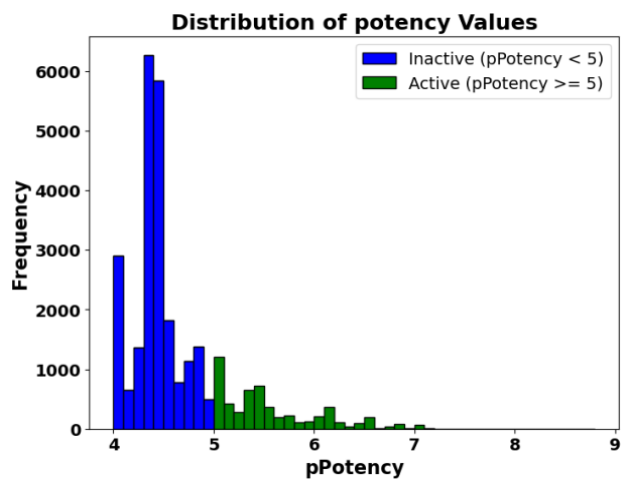

Fig. 2. Distribution of pPotency values for active (pPotency ≥ 5) and inactive (pPotency< 5) molecules in the dataset after data cleaning. The dataset contains 22688 inactive and 5569 active molecules, with a clear concentration of inactive compounds around pPotency = 4.5 and active compounds between pPotency 5 and 7.

### B. Performance of the Model

The trained neural network model demonstrated high classification accuracy in distinguishing active and inactive compounds. After 75 epochs of training, the model achieved a loss of 0.062, accuracy of 97.77%, and 94.24% for train and test data, respectively, indicating strong generalization to unseen data. The Receiver Operating Characteristic – Area Under the Curve (ROC-AUC) score was 0.94, further validating the model's capability to distinguish between the two classes. Table 1 shows the classification report. The model exhibited high precision (0.97) for inactive compounds and 0.91 for active compounds, ensuring a low false-positive rate. It also maintained a recall of 0.98 for the active class, effectively identifying the most active compounds.

TABLE I. CLASSIFICATION REPORT

| Class | Precision | Recall | F1-Score | Support |
|---|---|---|---|---|
| 0 (Inactive) | 0.97 | 0.91 | 0.94 | 4538 |
| 1 (Active) | 0.91 | 0.98 | 0.94 | 4538 |

### C. Docking of Predicted Molecules

Out of the 10,290 new molecules tested, the trained model predicted 1,092 as active. These inhibitors were then subjected to PAINS (Pan Assay Interference Compounds) and BRENK filtering, reducing the dataset to 396 molecules by eliminating false positives and compounds likely to interfere with biological assays [7, 8]. The binding affinity of the known inhibitor was recorded at -10.79 ± 1.45, which closely matched binding affinity values obtained for a few predicted inhibitors against SmTGR (Table II). Binding affinity quantifies the strength of interaction between a ligand and a protein. It is expressed as the Gibbs free energy of binding (ΔG), where more negative values indicate stronger interactions.

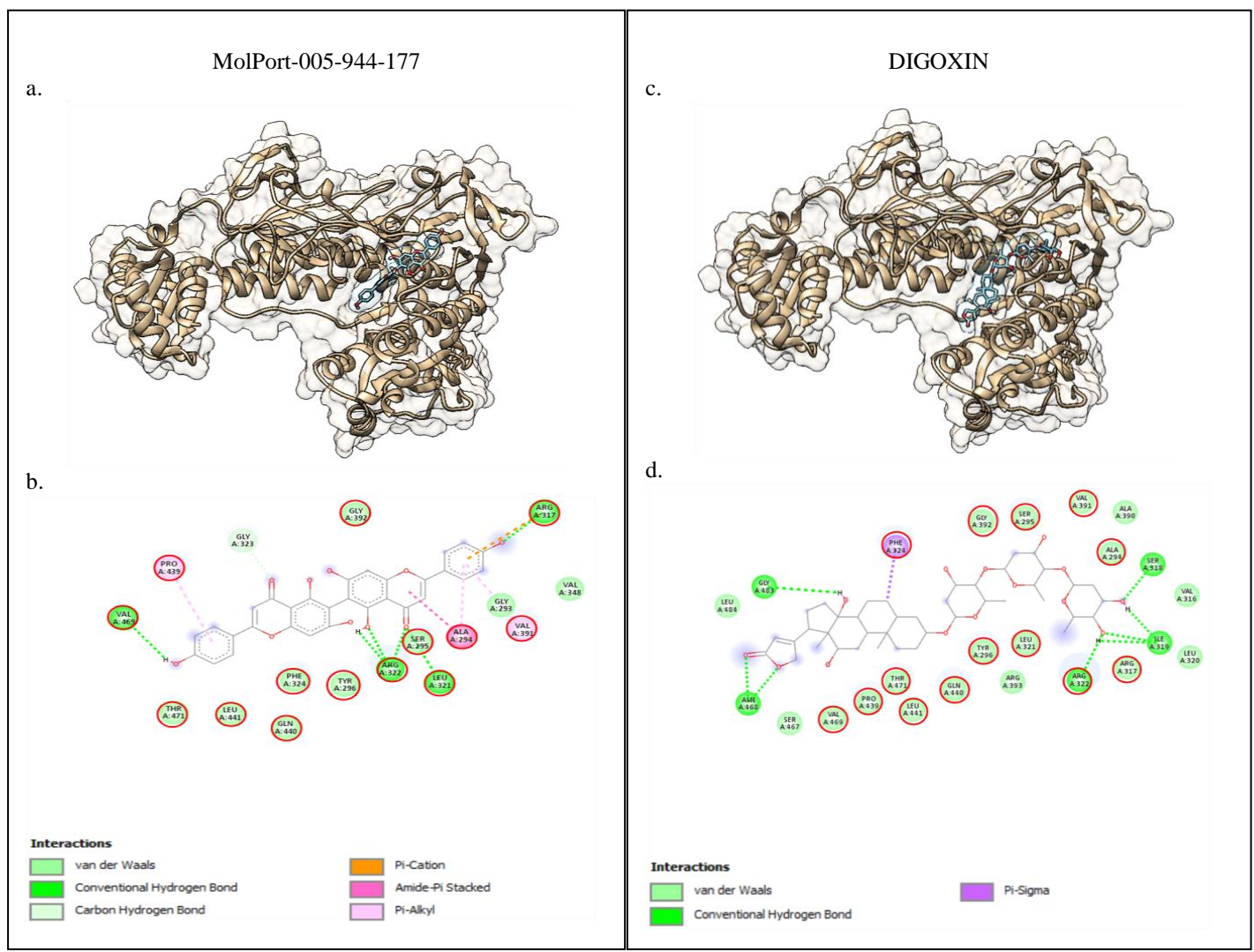

Fig. 3. Visual representation of docked complexes. (a) 3D-visualisation of MolPort-005-944-177, (b) 2D visualization of MolPort-005-944-177, (c) 3D-visualisation of Digoxin, (d) 2D visualization of Digoxin.

The binding affinity results suggested that the predicted inhibitors exhibited a similar inhibitory effect to Digoxin, indicating comparable efficacy due to their shared binding pockets. For example, MolPort-005-944-177 demonstrated a similar binding affinity of -10.39 ± 0.01 kcal/mol. The minimal standard deviation confirmed the reproducibility of the docking protocol. From the 3D visualization, it was evident that both Digoxin and the predicted inhibitor occupied the same binding pockets (Fig. 3(a), (b), (c)), suggesting their interaction with the active site of SmTGR.

TABLE II. BINDING AFFINITY VALUES OF TGR WITH TOP 6 PREDICTED INHIBITORS

| Sl. no | Prediction Probability | MOLPORT ID | Binding affinity (kcal/mol) |
|---|---|---|---|
| 1 | 0.99 | MolPort-039-338-190 | -10.76±0.03 |
| 2 | 0.98 | MolPort-044-754-024 | -10.44±0.30 |
| 3 | 0.81 | MolPort-005-944-491 | -10.19±0.69 |
| 4 | 0.95 | MolPort-005-944-177 | -10.39±0.01 |
| 5 | 0.95 | MolPort-039-052-672 | -10.25±0.01 |
| 6 | 0.99 | MolPort-035-705-947 | -10.18±0.01 |

Furthermore, the 2D visualization revealed that the docked complexes of Digoxin and MolPort-005-944-177 shared key binding residues—GLY 392, ARG 322, THR 471, GLN 440, VAL 391, PRO 439, PHE 324, TYR 296, SER 295, ALA 294, VAL 469, LEU 441, LEU 321, and ARG 317—as highlighted in red circles (Fig. 3). This overlap in binding interactions suggests that MolPort-005-944-177 may share a similar inhibitory mechanism with Digoxin against TGR. This further supports that the predicted inhibitors bind at the same site as the known inhibitor. A comparison of the ADMET (Absorption, Distribution, Metabolism, Excretion, and Toxicity) results indicated that while both Digoxin and MolPort-005-944-177 demonstrated favorable pharmacokinetic properties, Digoxin exhibited significantly higher toxicity [9]. This increased toxicity could lead to adverse effects, limiting its therapeutic potential. In contrast, MolPort-005-944-177 displayed a better safety profile, suggesting it may serve as a more viable and biocompatible drug candidate for TGR inhibition while maintaining effective bioactivity.

## IV. CONCLUSION

Schistosomiasis, one of the world's most neglected tropical diseases, is caused by trematodes (flukes) of the *Schistosoma* genus and ranks as the second most severe parasitic tropical disease after malaria. Developing inhibitors through wet lab studies is both time-consuming and costly, highlighting the need for faster and more efficient approaches. SmTGR is a protein that plays a key role in immune suppression during parasite invasion, enabling the parasite to complete its life cycle within the human host. Currently, Praziquantel is the only commercially available drug targeting this disease. This study successfully applied QSAR modeling and molecular docking to identify six potential SmTGR inhibitors. The QSAR model demonstrated robustness across diverse chemical profiles and was trained using a dataset from ChEMBL. Leveraging this trained model, active compounds were identified from the MolPort database. Furthermore, these molecules shared key binding residues with the known inhibitor Digoxin, confirming their interaction at the same binding site. In conclusion, QSAR modeling and molecular docking proved effective in identifying naturally derived, low-toxicity SmTGR inhibitors, offering promising candidates for further experimental validation.

## ACKNOWLEDGMENT

The authors thank AIC-DSU for supporting them with the server for conducting this study. The authors also thank KSCST for funding this work.